\documentclass[ALICE,manyauthors]{cernphprep}
\usepackage{cernphprep}
\usepackage{orcidlink}

\begin{document}

\begin{titlepage}
% the dates below correspond to CERN approval
% please don't touch: EB chairs will take care
\PHyear{2022}       % required, will be obtained from CERN
\PHnumber{225}      % required, will be obtained from CERN
\PHdate{25 October}  % required, will be obtained from CERN
%%%%%%%%%%%%%%%%%%%%%%%%%%%%%%%%%%%%%%%%%%%%%%%%%%%%

%%% Put your own title + short title here:
\title{\LARGE{Jet-like correlations with respect to \kzero and \lmb(\almb) in \pp \\
and central \PbPb collisions at $\mathbf{\it\sqrt{s_\mathrm{NN}}}$ = 5.02 TeV}}
\ShortTitle{Jet-like correlations with respect to \kzero and \lmb(\almb) in \pp and \PbPb collisions}   % appears on left page headers

%%% Do not change the next lines
\Collaboration{ALICE Collaboration\thanks{See Appendix~\ref{app:collab} for the list of collaboration members}}
\ShortAuthor{ALICE Collaboration} % appears on right page headers, do not change

\begin{abstract}

Two-particle correlations with \kzero, \lmb/\almb, and charged hadrons as trigger particles in the transverse momentum range $8 < p_\mathrm{T,trig}<16$ \GeVc, and associated charged particles within $1 < p_\mathrm{T,assoc}<8$ \GeVc, are studied at midrapidity in \pp and central \PbPb collisions at a centre-of-mass energy per nucleon--nucleon collision \fivenn with the ALICE detector at the LHC.
 After subtracting the contributions of the flow background, the per-trigger yields are extracted on both the near and away sides, and the ratio in \PbPb collisions with respect to \pp collisions (\iaa) is computed. The per-trigger yield in \PbPb collisions on the away side is strongly suppressed to the level of \iaa $\approx 0.6$ for $p_\mathrm{T,assoc}>3$ \GeVc as expected from strong in-medium energy loss, while an enhancement develops at low $p_\mathrm{T,assoc}$ on both the near and away sides, reaching \iaa $\approx 1.8$ and $2.7$ respectively. These findings are in good agreement with previous ALICE measurements from two-particle correlations triggered by neutral pions ($\pi^{0}$--h) and charged hadrons (h--h) in \PbPb collisions at \twosevensixnn. Moreover, the correlations with \kzero mesons and \lmb/\almb baryons as trigger particles are compared to those of inclusive charged hadrons. The results are compared with the predictions of Monte Carlo models.

\end{abstract}
\end{titlepage}
%%%%%%%%%%%%%%%%%%%%%%%%%%%%%%%%%%%%%%%%%%%%%%%%%%%%
\setcounter{page}{2} %please do not remove this line

%%%%%%%%%%%%%%%%%%%%%%%%%%%%%%%%
% begin main text
%%%%%%%%%%%%%%%%%%%%%%%%%%%%%%%%

\section{Introduction}%
\label{sec:intro}

Ultra-relativistic heavy-ion collisions produce the quark–gluon plasma (QGP), a high-temperature and high-energy-density state of strongly-interacting matter, which is characterised by the deconfinement of the colour charges, quarks and gluons ~\cite{BRAHMS:2004adc,PHOBOS:2004zne,STAR:2005gfr,PHENIX:2004vcz}. One of the most remarkable signatures of QGP formation is “jet quenching”, a modification of the parton showers which is induced by the interactions of energetic quarks and gluons with the medium. One of the manifestations of jet quenching is a reduction of the energy and transverse momentum ($p_\mathrm{T}$) of the jet (and of the hadrons emerging from jet fragmentation) due to medium-induced gluon radiation and collisions with medium constituents~\cite{1994,Jet_quenching1}, usually dubbed as parton energy loss.
It is generally measured through the suppression of jets and high-\pt hadron production, and the modification of angular correlations of these hadrons in heavy-ion collisions. The suppression is usually quantified via the nuclear modification factor, $\RAA$, which is defined as the ratio of the particle yield in \PbPb collisions to that in \pp collisions scaled by the average number of binary nucleon–nucleon collisions~\cite{ALICE:2010yje}. 

Jet quenching was first observed at the Relativistic Heavy Ion Collider (RHIC) in Au–Au collisions at centre-of-mass energies per nucleon--nucleon collision ${\sqrtsNN}$ = 130 and 200 \GeV~\cite{Suppression_PHENIX,STAR:2013thw}, and then at the Large Hadron Collider (LHC)~\cite{ALICE:2010yje,Jet_Suppression_CMS}. At LHC energies, jet quenching can be studied through fully reconstructed jets in heavy-ion collisions at high transverse momentum by measurements of inclusive jet $\RAA$~\cite{jet_suppression_ATLAS,ALICE:2015mjv,ALICE:2019qyj,ALICE:2015mdb} and dijet imbalance~\cite{ATLAS:2010isq,CMS:2012ulu,jet_quenching_CMS,CMS:2021nhn,ATLAS:2022zbu}, as well as through charged hadron $\RAA$~\cite{Suppression:2,Dependence_CMS,charged_particle_suppression_CMS,jet_fragmentation_CMS,ALICE:2012ab,ALICE:2012mj,RAA0,RAA1}. At low transverse momentum of less than 10~\GeVc, uncorrelated background fluctuations due to the underlying event dominate, and event-by-event jet reconstruction becomes difficult. Hence, di-hadron correlations provide an alternative way to study jets in this \pt regime.

Jet suppression has been studied by the ALICE Collaboration at the LHC via the two-particle correlation method~\cite{modification:SM,Jetlike_pion:SM} with neutral and charged hadrons as trigger particles in \PbPb collisions at \twosevensixnn. Both studies indicate a strong suppression of the per-trigger yield at high associated particle transverse momenta $p_\mathrm{T,assoc}$ of the away-side peak at an azimuthal angle difference  $\dphi = \pi$ in central \PbPb collisions compared to those in \pp collisions. 
Moreover, a strong  enhancement has been observed at low $p_\mathrm{T,assoc}$ of the near-side peak at $\dphi$ = 0.
The CMS Collaboration has drawn qualitatively similar conclusions~\cite{CMS_very_high}. These effects indicate the presence of a high-opacity
 medium produced in heavy-ion collisions with which the partons interact along their path.

Jet properties are expected to depend strongly on the species (quark/gluon) and the mass of the parton that initiated the jet, due to the difference in their fragmentation pattern. Differences between quark and gluon jets were observed in e$^{+}$e$^{-}$ annihilations at LEP. A fragmenting gluon emits more soft gluons than a fragmenting quark. Therefore, a gluon-initiated jet typically contains more particles and is characterised by a wider jet cone, while a quark jet is more collimated~\cite{LEP1,LEP2,Blanco:2021usa,OPAL}.
Furthermore, the fragmentation properties of different parton species in the hot-dense medium created in a heavy-ion collision are also expected to be different for quark- and gluon-initiated jets. Because of its larger colour charge, a gluon jet is expected to lose more energy than a quark jet~\cite{Qiu:2019sfj}. The quark and gluon jet fractions are thus expected to be modified in heavy-ion collisions with respect to pp collisions. The relative hadron production rates in quark and gluon jets seem to differ for \kzero and \lmb, with baryons produced more copiously in gluon jets~\cite{OPAL,Liu:2006sf}.   Therefore, the angular correlations involving these hadrons may shed some light on jet quenching effects for jets originating from quark and gluon initiated parton showers, whose interaction amplitude with the medium may differ.

In this article, the measurements of two-particle correlations are presented with \kzero, \lmb/\almb, and charged hadrons as trigger particles with transverse momentum $8 < p_\mathrm{T,trig}<16$ \GeVc, and associated charged particles with $1 < p_\mathrm{T,assoc}<8$ \GeVc, studying their angular difference, $\dphi$, distributions in \pp and most central \PbPb collisions at \fivenn. 
These measurements extend the previous ALICE studies of in-medium jet energy loss via angular correlations ~\cite{modification:SM,Jetlike_pion:SM,Multiplicitydependence,ALICE:2016clc,ALICE:2019jt,ALICE:2019sqi} by considering lower $p_\mathrm{T,assoc}$ charged hadrons ($p_\mathrm{T,assoc}$< 3 \GeVc), and by comparing between meson and baryon as trigger particles including strangeness.  These measurements have been extensively used to search for remnants of the radiated energy and the medium response to the high-$\pt$ parton, and further explore possible changes to parton fragmentation patterns and hadronisation mechanisms in the transition from the QGP to a hadron gas.
 
 This article further explores jet quenching effects by studying the nuclear modification factor (\iaa), which is defined as the ratio of the jet-like yield from nucleus--nucleus (A--A) collisions to the one from \pp collisions.   
 It provides information on the interplay between the parton production spectrum and the energy loss in the medium. 
 On the near side, a deviation of \iaa from unity can be connected to a change of the fragmentation function, a change of the quark versus gluon jet ratio, and a bias on the parton \pt spectrum. 
 On the away side, the suppression at high $p_\mathrm{T,assoc}$ (\iaa $<$1) is evidence of parton energy loss, while the enhancement at low $p_\mathrm{T,assoc}$ (\iaa $>$1) may involve an interplay of various contributions, such as $k_{\rm T}$ broadening, medium excitation, as well as fragments from radiated gluons~\cite{modification:SM,Jetlike_pion:SM,Large_angle}.

The article is organised as follows: Sec.~\ref{section2} provides details on the experimental setup, Sec.~\ref{section3}
describes the data analysis, Sec.~\ref{section4} is devoted to discussing the results, and Sec.~\ref{section5} contains the summary.

\section{Experimental setup}%
\label{section2}

The ALICE experiment is a multi-purpose detector designed to study hot and dense strongly-interacting
matter by measuring particles produced in ultra-relativistic heavy-ion collisions at the LHC. A detailed description of the ALICE
detector and its performance can be found in~\cite{ALICE:2014sbx,ALICE:JINST3}. The apparatus consists of a central barrel with several detectors covering the central rapidity region and positioned inside a solenoidal magnet providing a 0.5 T field parallel to the beam direction and a set of detectors placed at forward and backward rapidities. 
The detectors used for the present analysis are briefly described in this section.
In the central barrel, charged particles are reconstructed using the Inner Tracking System (ITS) and the Time Projection Chamber (TPC), which provide full azimuthal coverage for track reconstruction within a pseudorapidity range of $|\eta| < 0.9$. 
The ITS is designed to provide tracking near the primary vertex, and is thus a key detector for improving the primary vertex resolution and the reconstruction of secondary vertices from weak decays.
The ITS is comprised of six cylindrical layers of silicon detectors surrounding the beam pipe, which make use of three different detector technologies. In the four outer layers, Silicon Drift Detectors (SDD) and Silicon Strip Detectors (SSD) are used for tracking and particle identification. The two innermost layers are equipped with Silicon Pixel Detectors (SPD) covering the intervals $|\eta| < 2$ and $|\eta| < 1.4$, respectively. They provide high-resolution space points for the determination of the primary vertex position and of the track parameters in the vicinity of the beam axis.
The TPC is the main tracking detector of the ALICE central barrel. It is a large (88~$\rm m^3$) cylindrical detector filled with gas. The readout planes, located at the two edges of the drift volume, are composed of 18 azimuthal sectors that cover the full azimuth. Each sector contains 159 pad rows arranged radially. In addition, it performs particle identification (PID) via the measurement of the particle specific energy loss (d$E/\mathrm{d}x$) in the
detector gas. Additional information for particle identification is provided by the Time Of Flight (TOF) detector via the measurement of the charged-particle flight time from the interaction point to the detector.

The collision centrality is determined with the V0 detector.
The V0 system consists of two scintillator arrays located at asymmetric positions at both sides of the interaction point, one in the pseudorapidity range $2.8 ~<~ \eta ~<~ 5.1$  (\VZEROA) and the other at $-3.7 ~<~ \eta ~<~ -1.7$ (\VZEROC),
each made of four radial rings with each ring divided into eight sections in the azimuthal direction. 
It provides a signal proportional to the number of charged particles falling in these acceptance regions. The coincidence of signals in the V0A and V0C is also used to define the minimum-bias interaction trigger.

\section{Data analysis }
\label{section3}

The data sample used in this analysis consists of \PbPb collisions at \fivenn recorded by the ALICE detector in the 2015 and 2018 LHC data-taking campaigns and \pp collisions at \five collected in the year of 2017.

Accepted events are required to match a minimum-bias trigger, with a reconstructed vertex whose distance from the nominal interaction point along the beam direction is within 7 cm and 10 cm in \PbPb and \pp collisions, respectively.
These selection criteria were met by approximately 70 million events in the 0–10\% centrality interval in \PbPb collisions, corresponding to the 10\% most central (highest-multiplicity) collisions, and by approximately a billion \pp collisions.

The contamination from beam-induced background is removed offline by using the timing information in the V0 detectors and taking into account the correlation between tracklets and clusters with the SPD detector, as discussed in detail in ~\cite{ALICE:2014sbx}. The contamination from in-bunch pile-up events is removed offline excluding events with multiple vertices reconstructed with the SPD.

The centrality estimation in \PbPb collisions is based on the sum of the amplitudes from the two V0 detectors which is proportional to the number of charged particles produced in the V0 acceptance.

The analysis consists of three main parts: 
(i) \kzero, \lmb(\almb), and primary charged particle selection; (ii) construction of the correlation function and of the corresponding corrections; (iii) systematic uncertainty estimation.

\subsection{\texorpdfstring{\kzero, \lmb(\almb)}{Lg} and primary charged particle selection}
\label{selection}
The analysis procedure begins with the reconstruction of the trigger and associated particles defined in Sec.~\ref{sec:intro}.
% introduction (L45-47).
The identification and reconstruction of \kzero and \lmb(\almb) particles follows the approach used in previous measurements presented in~\cite{K0sandLambda_PbPb,hadron_correlations_pp,Anisotropic_flow}, via their most probable decay channels~\cite{ALICE:2014sbx} and by exploiting their characteristic \Vdecay decay topologies:
\[\kzero\rightarrow \pi^- + \pi^+, \quad 
\mathrm{BR}= 69.2\%,  \]
\[\lmb\rightarrow \pi^- + \rm{p}, \quad  \quad  \mathrm{BR}= 63.9\% ,\]
\[\almb\rightarrow \pi^+ + \rm{\overline{p}}, \quad \quad  \mathrm{BR}= 63.9\%,\]
where BR denotes the branching ratio~\cite{PhysRevD.98.030001}.

Primary charged particles are reconstructed in the pseudorapidity range $|\eta| < 0.8$ by combining the information from the ITS and TPC detectors (see Sec.~\ref{section2}). To maximise the tracking efficiency and improve the specific energy loss (d$E/\mathrm{d}x$) and momentum resolution for primary charged particles, and guarantee an optimal PID quality, tracks reconstructed with both the TPC and the ITS are required to cross at least 70 TPC readout rows out of a maximum of 159. The ratio of the number of crossed pad-rows to the number of findable clusters (that is the number of geometrically possible clusters which can be assigned to a track) is restricted to be greater than 0.8 (see ~\cite{ALICE:2014sbx} for the details). The goodness-of-fit values $\chi^{2}$ per cluster ($\chi^{2}/N_{\mathrm{clusters}}$) of the track fit in the TPC must be less than 2.5. Furthermore, in order to suppress the contamination of secondary particles from weak decays and interactions in the detector material, and ensure uniform distributions in the $(\eta,\phi)$ plane, additional selections are applied, leading to the sample of so-called "hybrid" tracks. These are composed of two independent samples of tracks selected with different criteria. The first sample consists of tracks with at least one hit in either of the two SPD layers and small distance of closest approach to the primary vertex. To compensate for the acceptance inhomogeneities introduced by the SPD dead areas and recover good momentum resolution, they are combined with a second sample, which is constituted of tracks without any hit in the two SPD layers but constrained to the primary vertex~\cite{ALICE:2019wqv,ALICE:2020siw}. 
Another tracking mode that relies on the combination of the TPC and the ITS detectors, called global tracking, with tighter selection criteria in addition to requirements for clusters in the SPD or the SDD detectors was used for systematic checks in this analysis.

The identification of \Vdecay candidates is based on the geometrical selections on the decay topology reported in Table~\ref{tabel:Topological_selction} within $\abs\eta <0.7$ and on the particle identification of their decay products via the measured specific energy loss d$E/\mathrm{d}x$ in the TPC. Only \Vdecay candidates that had daughter tracks within 3$\sigma$ from the expected energy loss for the correct particle species were used, where $\sigma$ is the TPC d$E/\mathrm{d}x$ resolution (for more details see ~\cite{ALICE:2014sbx}).
The pairs of identified tracks are combined into \Vdecay candidates, and are accepted if their invariant mass $m_{\Vdecay}$ is within $\pm 3\sigma$ from the nominal value of the \kzero, \lmb(\almb) mass~\cite{PhysRevD.98.030001}.
Because of the large combinatorial background in \PbPb collisions, a number of topological selections had to be more restrictive than those used in the \pp analysis. In particular, more stringent selections were applied on the distance of closest approach (DCA) of the decay tracks to the primary vertex (PV) and on the \Vdecay radius, which is the distance between the point where the \Vdecay decays (secondary vertex) and the PV.

To suppress the contamination of \lmb reconstructed as \kzero, a selection is applied on the Armenteros-Podolanski (Arm-Pod) variables, taking the form $\pt^{\mathrm{arm}}~>~0.2|\alpha^{\mathrm{arm}}|$, where $\pt^{\mathrm{arm}}$ is the transverse component of the daughter track momentum in the \Vdecay momentum system and $\alpha^{\mathrm{arm}}$ is defined as $\alpha^{\mathrm{arm}}=(p^{+}_{||}-p^{-}_{||})/(p^{+}_{||}+p^{-}_{||})$, where  $p^{+}_{||}(p^{-}_{||})$ is the projection of the positively (negatively) charged decay product momentum on the momentum of the \Vdecay~\cite{Podolanski_var1,Podolanski_var2}.
Additional selections are applied to remove the \Vdecay decays reconstructed in the TPC, which are produced in pile-up collisions occurring in different bunch crossings within the TPC readout time. These out-of-bunch pileup events are not rejected by the event selection criteria described above. The criterion for this removal is that at least one of the \Vdecay daughter tracks is reconstructed in both the ITS and the TPC, or has a signal in the Time-Of-Flight detector.

\begin{table}[h]
\centering
    \caption{The selection criteria for \Vdecay candidates. The angle $\theta_{\rm PA}$ is defined by the \Vdecay momentum and the line connecting the PV and the \Vdecay decay point. The proper lifetime $mL/p$ is calculated with $m$ as the expected \Vdecay mass under the current hypothesis, $L$ is the distance between the primary and secondary vertices, and $p$ is the particle momentum.}

\begin{tabular}{c|c|c} \hline \hline

 \textbf{ \kzero selection criteria}                     & \textbf{\PbPb}        & \textbf{\pp}\\ \hline \hline
\hspace{-0.2cm}   DCA of daughter to PV (cm)                                & $>$ 0.1                & $>$ 0.06 \\
   DCA between daughters ($n_{\rm \sigma}$)                                & $<$ 0.8                 & $<$ 1  \\
   cos($\theta_{\rm PA}$)                            & $>$ 0.97                   & $>$ 0.97\\
   \Vdecay radius (cm)                                      & $>$ 5                    & $>$ 0.5  \\
\hspace{-0.5cm}   Proper lifetime $mL/p$ (cm)                               & $<$ 20                    &$<$ 20 \\ \hline \hline
   
   \textbf{\lmb(\almb) selection criteria}  & \textbf{\PbPb}         & \textbf{\pp}\\  \hline \hline
\hspace{-0.5cm}   DCA of proton to PV (cm)                                & $>$ 0.1                   & $>$ 0.06   \\
\hspace{-0.9cm}   DCA of pion to PV (cm)                                   & $>$ 0.25                    &$>$ 0.06 \\
  DCA between daughters ($n_{\rm \sigma}$)  & $<$ 0.8                 & $<$ 1 
  \\
 cos($\theta_{\rm PA}$)                           & $>$ 0.995                   & $>$ 0.98\\
   \Vdecay radius (cm)                                    & $>$ 5                    & $>$ 0.5  \\
\hspace{-0.5cm}   Proper lifetime $mL/p$ (cm)                              & $<$ 25                      & $<$ 30 \\
   \hline
\end{tabular}
	\label{tabel:Topological_selction}
\end{table}

\vspace{0.7cm}
\subsection{Correlation function and corresponding corrections}
\label{correlation and corr}
Unidentified charged particles with transverse momentum within $1 < p_\mathrm{T,assoc}<8$ \GeVc associated to \Vdecay (\kzero, \lmb/\almb) and charged hadrons with transverse momentum $8 < p_\mathrm{T,trig}<16$ \GeVc are studied via two-particle angular correlations where $p_\mathrm{T,assoc}$ < $p_\mathrm{T,trig}$. The correlation function is defined as a function of the azimuthal angle difference $\dphi = \varphi_{\rm{trig}} - \varphi_{\rm{assoc}}$ and pseudorapidity difference  $\deta = \eta_{\rm{trig}}- \eta_{\rm{assoc}}$ between trigger and associated particles. It is expressed as

  % \vskip 0.2in
\begin{align}
  C(\dphi, \deta)&=\frac{1}{N_{\rm{trig}}}\frac{\mathrm{d}^2N_{\rm{assoc}}}{\mathrm{d}\Delta\varphi \mathrm{d}\Delta\eta},
    \label{correlation 2D}
\end{align}
    
\noindent  where {\Ntrig} is the number of trigger particles and {\Nass} is the number of associated particles~\cite{K0sandLambda_PbPb,Multiplicitydependence,Long_range_correlations_PPb}.
Each trigger and each associated particle is weighted with a correction factor that accounts for reconstruction efficiency and contamination by secondary particles. These corrections are
applied as a function of $\eta$ and \pt.
For the evaluation of the charged-particle reconstruction efficiency, \pp collisions are simulated with the Monte Carlo (MC) generator PYTHIA8.125~\cite{PYTHIA1} (Monash 2013 tune~\cite{PYTHIA2}), while \PbPb collisions are simulated using the HIJING event generator~\cite{HIJING_model}. The generated particles are transported through the apparatus, which is modelled in the simulation using the GEANT3 transport code~\cite{Brun:1994aa}.

Pair acceptance corrections are evaluated with a mixed-event technique, where $(\Delta\varphi, \Delta\eta)$ distributions with the trigger and the associated particles originating from different events with similar multiplicity and PV position along the beam axis are constructed. The mixed-event correlation is scaled to unity with a factor equal to the average of $\Delta\varphi$ bins with $\Delta\eta=0$.

In the analysis reported in this paper, a small relative rise in the correlation structure around $\Delta\varphi=\pi$ at large $\Delta\eta$ has been observed. This structure, which was already reported in previous studies~\cite{PHOBOS:2007whv,ALICE:2016mfw,hadron_correlations_pp}, is usually indicated as “wings”. These wings are an artifact introduced by the mixed-event correlation, which does not match the shape of the background perfectly due to the finite binning in multiplicity and PV position along the beam axis. Hence, the correlation function is scaled once more with a 2D distribution constant in $\Delta\varphi$ and dependent on $\Delta\eta$ in order to get a flat distribution in $\Delta\eta$ in the away side.

In the case of \Vdecay trigger particles, the correlation function in Eq.~\ref{correlation 2D} is obtained for \kzero and \lmb candidates reconstructed in the invariant mass ranges $|m_{\Vdecay}-M|<3\sigma$, $M$ being the corresponding mass of the particle as given by the PDG~\cite{PhysRevD.98.030001}.  
However, the angular correlation also contains contributions from combinatorial background candidates in this mass range, and this source of background needs to be subtracted. 
This contribution is estimated via the per-trigger correlation distribution built from background trigger candidates in the side-band invariant mass ranges $5\sigma<|m_{\Vdecay}-M|<8\sigma$.

For the measurement of \iaa, the yield of associated particles per trigger particle is studied as a function of the azimuthal angle difference $\dphi$ in \pp and \PbPb collisions. This distribution is given by

 \begin{align}
 C(\Delta\varphi)&=\frac{1}{N_{\rm{trig}}}\frac{\mathrm{d}N_{\rm{assoc}}}{\mathrm{d}\Delta\varphi } .
  \centering 
    \end{align}
    This quantity is measured for all pairs of particles where $p_\mathrm{T,assoc}< p_\mathrm{T,trig}$   within $|\deta|< 0.7$ as a function of $p_\mathrm{T,assoc}$. Since the anisotropic flow induces a $\dphi$ dependent modulation of the background that cannot be neglected in \PbPb collisions, the background contributions to the yield extraction in \PbPb are modelled as 
     
\setlength{\abovedisplayskip}{0pt}
\setlength{\belowdisplayskip}{0pt}

      \begin{equation}
     B(\Delta\varphi)=B_{0}\left(1+2\sum_{n} v_{n}^{\rm{trig}}v_{n}^{\rm{assoc}}\cos(n\Delta\varphi)\right),
         \label{Eq:Bkg} \end{equation}

 where {\vntrig} and  {\vnass} are the flow coefficients for trigger and associated particles, respectively. Note that in this analysis only values of $n = 2,3$ are considered.
 The data of {\vnass} for charged particles and {\vntrig} for \Vdecay particles are taken from Ref.~\cite{ALICE:2018yph}.

 The constant $B_{0}$ is determined with the zero yield at minimum (ZYAM) method, i.e., it is defined as the average value of six bins in the $C(\Delta\varphi)$ distributions outside of the jet peaks where the background is assumed to be flat~\cite{Anisotropic_flow,Long_range_correlations_PPb,Jetlike_pion:SM,ZYAM_flow}. The ZYAM method is also used to subtract the pedestal in \pp collisions. Finally, the jet-like correlations are obtained as

  \begin{equation}
     J(\Delta\varphi) = C(\Delta\varphi)- B(\Delta\varphi).
 \end{equation}

The per-trigger jet-like charged hadron yield is extracted by integrating the jet-like correlation function over a given $\dphi$ range according to the following equation: 

  \begin{equation}
     Y_{\Delta\varphi} = \int_{\it X} J(\Delta\varphi)\rm{d}\Delta\varphi ,
 \end{equation}

where $X$ denotes either the near-side |$\dphi$| < 0.9 or the away-side |$\dphi-\pi|<1.2$ region to be integrated.

The nuclear modification  factor, \iaa, is then calculated by taking the ratio of the per-trigger jet-like yield in \PbPb collisions over that in pp collisions as

\begin{align}
I_\mathrm{AA} =\frac{Y^{\PbPb}_{\Delta\varphi}}{Y^{\rm{\pp}}_{\Delta\varphi}}.
\centering
\label{Eq.IAA}
\end{align}

\iaa is measured for the near and away sides in 0–10\% central \PbPb collisions at \fivenn, for unidentified charged hadrons with transverse momentum $1 < p_\mathrm{T,assoc}<8$ \GeVc associated with \kzero, \lmb/\almb, and charged hadron triggers with transverse momentum $8 < p_\mathrm{T,trig}<16$ \GeVc.

\vspace{-0.1cm}

 \subsection{Systematic uncertainty estimation} 
The sources of systematic uncertainties for the per-trigger yield and \iaa measurements on both the near and away sides in \PbPb and \pp collisions are listed in Tables~\ref{table_sys_K} and~\ref{table_sys_L} for \kzero--h and ($\Lambda$+\almb)--h angular correlations, respectively and in Table~\ref{table_sys_h} for h--h correlations. 
Since the uncertainties are $p_\mathrm{T,assoc}$ dependent, the minimum and maximum values of the contribution of each source are shown. 
The following main sources of systematic uncertainty are considered: systematic effect originating from the detector acceptance and, in particular, from the selection on the primary vertex position along the beam axis, systematic related to the track and \Vdecay reconstruction and selection, systematic due to the additional cut of Arm-Pod variables for \kzero candidates, systematic uncertainty induced by the applied corrections (background invariant mass range, binning and scaling of the event-mixing correlations, wing correlation shape at large $\deta$, pedestal determination), systematic uncertainty related to the near-side jet acceptance and by the $\dphi$ integration window.

For the uncertainty related to the detector acceptance, the primary vertex position selection along the beam axis is decreased from $\pm{10}$ cm to $\pm{8}$ cm from the nominal interaction point position in the case of \pp collisions and from $\pm{7}$ cm to $\pm{8}$ cm in the case of \PbPb collisions.

Systematic uncertainties related to the \Vdecay reconstruction are estimated by varying the topological selection criteria around their nominal values listed in Table~\ref{tabel:Topological_selction}. Two new sets, based on more and less restrictive selections, were chosen to increase or reduce the selected yield by about 10\%.  Each has a different magnitude and \pt dependence of the efficiency, thus impacting on the \pt distribution of selected \Vdecay trigger particles and on their correlation shapes.

The cut on Arm-Pod variables for \kzero candidates is varied for evaluated the systematic to a looser value of $\pt^{\mathrm{arm}}>0.17|\alpha^{\mathrm{arm}}|$ and a tighter value of $\pt^{\mathrm{arm}}~>~0.23|\alpha^{\mathrm{arm}}|$.\\
\vspace{0.04cm}
The uncertainty from track selection is estimated by varying the track selection criteria between hybrid 
\vspace{-0.09cm}
tracks and global tracks discussed in Sec.~\ref{selection}. In the case of global tracks, a stricter transverse momentum

dependent requirement in the value of the DCA in the transverse plane results in reducing even further the amount of secondary particles in the track sample.

In addition, the sideband invariant mass intervals used to estimate the combinatorial background under the \Vdecay signals (see Sec ~\ref{correlation and corr}) are varied within $\pm1\sigma$ in order to estimate the uncertainty related to the background subtraction procedure.

The event-mixing technique is used for the acceptance correction by combining tracks originating from events close in multiplicity and PV position. The vertex binning is then varied as well as the mixed-event scaling factor. Both variations produce non-negligible contributions to the total systematic uncertainties especially for \kzero and \lmb.
For more details see Tables ~\ref{table_sys_K},~\ref{table_sys_L},~\ref{table_sys_h}.

The wing effect at large $\deta$ which is only observed in \PbPb collisions is studied. The contribution to the systematic uncertainties is found to be small for the measurements of \lmb--h and h--h correlations while it is negligible for \kzero--h correlations.

The $\deta$ range is varied within 0.1 around its nominal value $\abs\deta$ < 0.7 in order to estimate the uncertainty related to the near-side jet acceptance.

The uncertainty from the pedestal determination is estimated by shifting the ZYAM determination from using the default six bins to four bins on the left and right side of the near-side peak in the ${\Delta\varphi}$ distribution, as well as by fitting the full ${\Delta\varphi}$ region between the near and away sides. 
The largest effect to the uncertainty on the per-trigger yields for \PbPb collisions arises from this source, especially at low \pt.

Finally, the integration window for the near and away side is varied around the nominal values ${|\Delta\varphi|} <0.9$ and ${|\Delta\varphi -\pi|} <1.2$ by $\pm$ 0.2 radians.

\vspace{-1mm}
The systematic uncertainties on the per-trigger yields in \pp and \PbPb collisions as well as on \iaa are computed for each source separately. Fluctuations from the varying statistics in the variation can
also affect the systematic of \iaa. Due to correlations, the systematic uncertainties from some of the sources (e.g. invariant mass range and track selection) cancel out partially in the \iaa. The resulting uncertainties depend on $p_\mathrm{T,assoc}$ and only the minimum and maximum values are reported in Tables~\ref{table_sys_K},~\ref{table_sys_L},~\ref{table_sys_h}. Finally, the total uncertainties are computed for the yields and the \iaa by summing in quadrature the contributions of the different sources.

%\vspace{0.2cm}

%\begin{center}
\begin{table}[h!]
\caption{Summary of the different sources and corresponding estimated systematic uncertainties for the per-trigger yield of \kzero--h correlations on the near and away sides in \pp and $\PbPb$ (0--10\%) collisions, as well as for the \iaa. The systematic uncertainties depend on $p_\mathrm{T,assoc}$ and for each source the minimum and maximum values are reported.}

\begin{tabular}[!ht]{|l|c|c|c|c|c|c|c|}\hline
%	\centering
		& \multicolumn{3}{c|}{\textbf{Near side (\kzero--h)}} & \multicolumn{3}{c|}{\textbf{Away side (\kzero--h)}}\\
		\hline
	\textbf{Uncertainty source} & $Y_{\dphi}^{\rm{\pp}}$ & $Y_{\dphi}^{\PbPb}$ & \iaa &$Y_{\dphi}^{\rm{\pp}}$ & $Y_{\dphi}^{\PbPb}$ & \iaa  \\	\hline
		
		Primary vertex $z$ &0--1.5\% & 4--5\% &4--7\% & 0--2.7\% &2--5\% &2--7\%  \\
		\hline
		\Vdecay topology selections &0--0.4\% &1--2\% & 0--1.8\% & 0--0.1\% &1--6\% &1--2\%  \\
		\hline

         Arm-Pod variables &--&0.9--2.3\% & 0.9--2.3\% &-- &0.6--2.5\% &0.6--2.5\%  \\
		\hline
  
		Track selection &0--1\% &1--2\% &1--4\% &0--1\% &2--4\% &2--5\%  \\
	    \hline
		Invariant mass range &0--2.8\% &1--2\% &1--2\% &0--2.7\% &2--3.5\% &0--1\%   \\
		\hline
		Mixing binning &0--0.4\% &0--2\% &1--2\% &0--1.7\% &0--6\% &2--8\%   \\
	    \hline
		Mixing scale & 0--1.5\% &1--3.5\% &1--6\% &0--2.7\% &1--6\% &1--7\%  \\
		\hline
		Wings correction  &-- &-- &-- &-- &-- &--   \\
		\hline
		$\deta$ range &0--1.4\% &0.5--2.3\% &0.4--1.6\% &--&-- &--\\
       \hline
		Pedestal determination &0--0.4\% &1--10\% &1--9\% &0--1\% &0--8\% &2--11\%  \\
	    \hline
  		Integration window  &0--0.2\% &1--3.6\% &1--3\% & 0--1\% &1--2\% &1--7\%  \\
		\hline
	  \textbf{Total} &0--3.8\% & 4.8--13.3\% &4.8--14.5\%&0--5.2\% &3.9--15.3\% &4.4--19.2\%   \\
		\hline
		\hline
	\end{tabular}
\label{table_sys_K}
\end{table}

%\vspace{4cm}

\begin{table}[h!]
	\caption{Summary of the different sources and corresponding estimated systematic uncertainties for the per-trigger yield of ($\Lambda$+\almb)--h correlations on the near and away sides in \pp and \PbPb (0--10\%) collisions, as well as for the \iaa. The systematic uncertainties depend on $p_\mathrm{T,assoc}$ and for each source the minimum and maximum values are reported.}

\begin{tabular}[h!]{|l|c|c|c|c|c|c|c|}\hline
		& \multicolumn{3}{c|}{\textbf{Near side ($\Lambda$+\almb)--h}} & \multicolumn{3}{c|}{\textbf{Away side ($\Lambda$+\almb)--h}}\\
		\hline
		\textbf{Uncertainty source} & $Y_{\dphi}^{\rm{\pp}}$ &$Y_{\dphi}^{\rm{\PbPb}}$ & $\iaa$ &$Y_{\dphi}^{\rm{\pp}}$ & $Y_{\dphi}^{\rm{\PbPb}}$& $\iaa$  \\	
		\hline
		
		Primary vertex $z$ &0--1.5\% &2--6\% &3--9\% &0--1.8\% &5--7\% &3--6\%  \\
		\hline
		\Vdecay topology selections &0--1.3\% &1--3.5\% &1--5\% &0--1\% &0--6\% &0--3\%  \\
		\hline
		
		Track selection &0--2.2\% &1--4.2\% &0--4.2\% &0--2.5\% &1--4\% &1--3\%\\
		
		\hline
		Invariant mass range&0--2.5\% &0--5\% &0--3\% &0--2.5\% &1--5\% &0--3\% 
		\\
		\hline
		Mixing binning &0--1.7\% &1--5\% &2--7\% &0--1.8\% &0--3.5\% &2--5\%
		\\
		\hline
		Mixing scale &0--2.8\% &1--3 \%& 2--10\% &0--1.7\% &0--4\% &2--11\%  \\
		\hline
		Wings correction  &-- &0--1.8\% &0--1.8\% &-- &0--1.8\% &0--1.8\%   \\
        \hline
		$\deta$ range &0--1.2\% &0--2.5\% &0--3\% &--&-- &--\\
		\hline
		Pedestal determination &0--0.2\% &3--11\% &2--15\% &0--0.4\% &1--9\% &2--7\%  \\
		\hline
		Integration window  &0--0.4\% & 1--3.5\% &1--4 \%&0--1 &1--5\% &1--5\%\\
	    \hline

	  \textbf{Total}  &0--5.2\% &4.2--16.4\% &4.8--23.1\% &0--4.9\% &5.4--16.2\% & 4.8--17\%  \\
		\hline
		\hline
	\end{tabular}
\label{table_sys_L}
\end{table}
\begin{table}[ht]
\caption{Summary of the different sources and corresponding estimated systematic uncertainties for the per-trigger yield of h--h correlations on the near and away sides in \pp and \PbPb (0--10\%) collisions, as well as for the \iaa. The systematic uncertainties depend on $p_\mathrm{T,assoc}$ and for each source the minimum and maximum values are reported.}
	\begin{tabular}[!ht]{|l|c|c|c|c|c|c|c|c|}\hline
		& \multicolumn{3}{c|}{\textbf{Near side (h--h)}} & \multicolumn{3}{c|}{\textbf{Away side (h--h)}}\\
		\hline
		\textbf{Uncertainty source} & $Y_{\dphi}^{\rm{\pp}}$& $Y_{\dphi}^{\rm{\PbPb}}$ & $\iaa$ &$Y_{\dphi}^{\rm{\pp}}$ & $Y_{\dphi}^{\rm{\PbPb}}$& $\iaa$  \\ 
		
		\hline
		Primary vertex $z$ &0--0.2\% &0--3\% &0--3\% &0--0.2\% &2--3\% &2--4\%  \\
		\hline
		Track selection  &0--1.7\% &0--1.5\% &0--1\% &0--1.1\% &1--2\% &0--3\% \\
		
		\hline
		Mixing binning &0--0.2\% &0--2\% &0--2\% &0--0.2\% &0--5.3\% &0--2\%  \\
		\hline
		Mixing scale &0--1.7\% &0--1.5\% &0.2--3.8\% &0--1.8\% &0--2\% &0--4\%\\
		\hline
		Wings correction &-- &0--1.2\% &0--1.2\% &-- &0--1.3\% &0--1.3\%  \\
		\hline
		$\deta$ range &0--0.7\% &0--2\% &0--1.6\% &--&-- &-- \\
		\hline
		Pedestal determination &0--0.2\% &0--5\% &0--4\% &0--0.2\% &2--7.7\% & 2--9.5\%  \\
		\hline
		Integration window  &0--0.2\% &0--0.7\% &0--2\% &0--0.2\% &0--1\% &0--1\%  \\
		\hline
	  \textbf{Total}  &0--2.5\% &0--7\% &0--7.2\% &0--2.2\% & 3--10.3\% &1.4--11.7\%\\
		\hline
		\hline
\end{tabular}
\label{table_sys_h}
\end{table}

\setlength{\textfloatsep}{1.5cm}  %% Or whatever length

%\vspace{cm}

\section{Results and discussion}%
\label{section4}

\subsection{Projection of the correlation function \texorpdfstring{$C(\Delta\varphi)$}{Lg}}
\begin{figure}[!hbt]
\includegraphics[width=18cm]{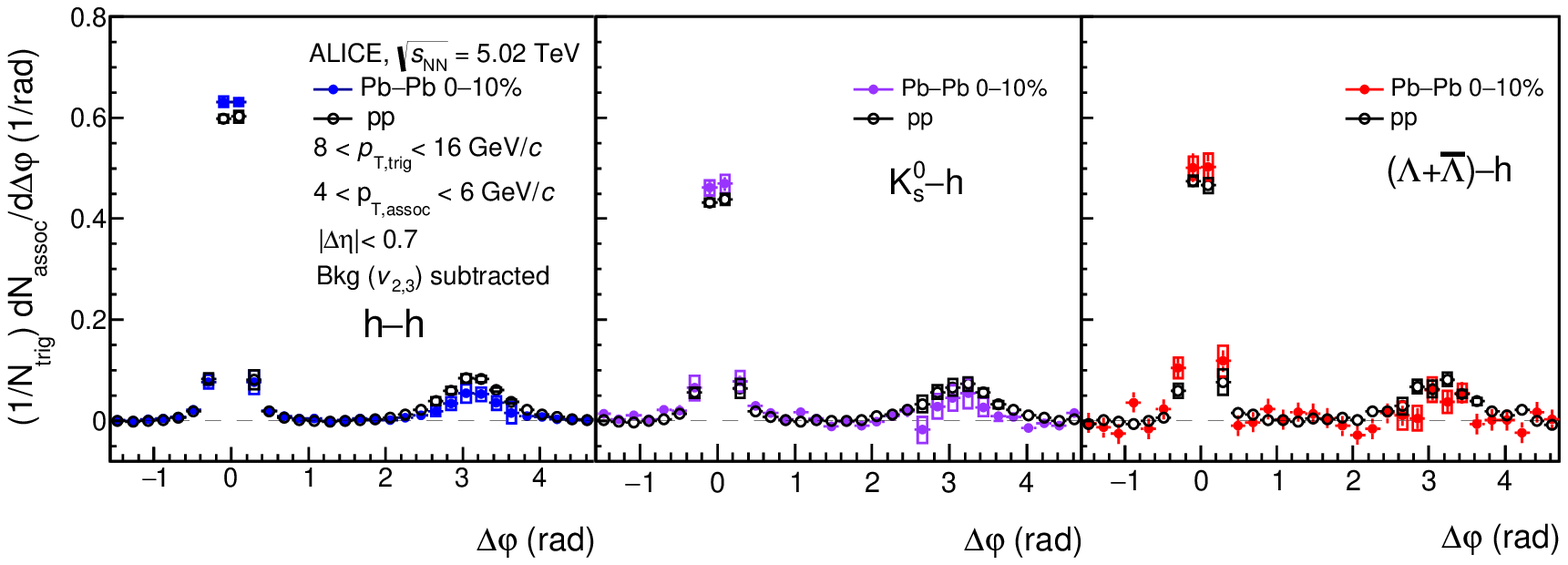}
\caption{Distributions of $C(\Delta\varphi)$ for h--h (left), \kzero--h (middle) and ($\Lambda$+\almb)--h (right) trigger particles with $8 < p_\mathrm{T,trig}<16$ \GeVc and associated particles with  $4 < p_\mathrm{T,assoc}<6$ \GeVc in 0--10\% central \PbPb (solid points) and \pp (open circles) collisions.
The error bars represent the statistical uncertainties while the boxes represent the systematic uncertainties. The background has been subtracted based on the estimation of ZYAM in \pp collisions and the additional contributions of the anisotropic flow harmonics $v_{2}$ and $v_{3}$ in \PbPb collisions.}
\label{fig:dphi}
\end{figure}

The distributions of $C(\Delta\varphi)$ for the three different trigger particles with transverse momentum within $8 < p_\mathrm{T,trig}<16$ \GeVc, and associated charged particles of $4 < p_\mathrm{T,assoc}<6$ \GeVc, are presented in Fig.~\ref{fig:dphi} for \pp and 0--10\% central  \PbPb collisions. A uniform background in the case of \pp is assumed, while the additional contributions of the anisotropic flow coefficients $v_{2}$ and $v_{3}$ in the case of \PbPb are subtracted.

The comparisons of $C(\Delta\varphi)$ distributions show a less pronounced away-side peak in central \PbPb collisions compared to \pp collisions for the three trigger particle species.
The per-trigger yields, obtained by integrating the correlation functions in the ranges of $|\dphi|<$0.9 on the near side and $|\dphi-\pi|<$1.2 on the away side have been studied as a function of $p_\mathrm{T,assoc}$.

\newpage

\subsection{Nuclear modification factor \texorpdfstring{\iaa}{Lg}}
\begin{figure}[!hbt]
\includegraphics[width=16cm]{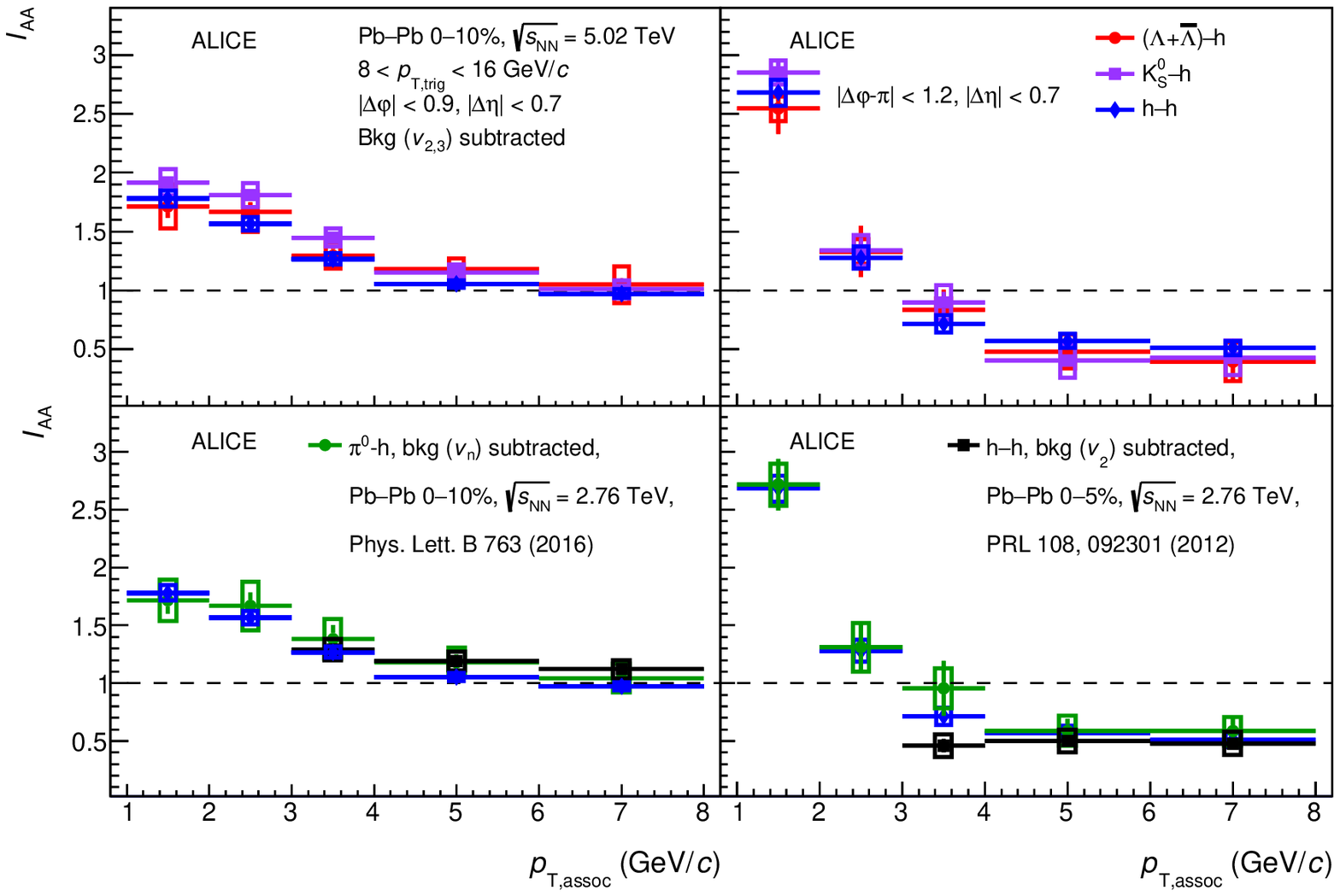}

\caption{Top panel: per-trigger yield modification, \iaa, of \kzero--h (square points), ($\Lambda$+\almb)--h (circle points) and h--h (diamond points), on the near side (left) and away side (right). Bottom panel: comparison of \iaa of h--h (diamond points) from this measurements at \fivenn, with \iaa of h--h (square points) and $\pi^{0}$--h (circle points) at \twosevensixnn, on the near side (left) and away side (right)~\cite{modification:SM,Jetlike_pion:SM}.}
\label{fig:all_IAA}
\end{figure}

The ratio of the integrated peak yields for h--h, \kzero--h, and ($\Lambda$+\almb)--h angular correlations in 0--10\% central \PbPb collisions relative to \pp collisions, \iaa defined in Eq.~\ref{Eq.IAA}, is shown in the upper panels of Fig.~\ref{fig:all_IAA} as a function of the associated particle \pt, $p_\mathrm{T,assoc}$.
 In the away-side \iaa (top-right panel), a suppression of per-trigger yields in central \PbPb collisions to the level of \iaa $\approx 0.6$ for $p_\mathrm{T,assoc}>3$ \GeVc is observed for all trigger particles.
This is expected to arise from the jet quenching effect on energetic partons caused by in-medium energy loss. However, going down in $p_\mathrm{T,assoc}$, the \iaa ratio turns from suppression to enhancement reaching \iaa $\approx 2.7$.
Such an enhancement could be expected based on energy conservation, with the high-\pt partons losing energy that gets converted into low-\pt associated particles.
On the near side (top-left panel), a moderate enhancement of per-trigger yields in central \PbPb collisions for $p_\mathrm{T,assoc}< 4$ \GeVc is observed, with \iaa reaching up to a value of about $\approx$ 1.8 for
1 < $p_\mathrm{T,assoc}<3$ \GeVc. This enhancement suggests that the near-side parton is also subject to medium effects~\cite{modification:SM}.

For associated particles with $p_\mathrm{T,assoc}$ > 4 \GeVc, the near-side \iaa results suggest a significantly smaller difference between \pp and \PbPb collisions, which is consistent with trigger particles (having $p_\mathrm{T,trig}$ > $p_\mathrm{T,assoc}$) predominantly coming from the vacuum-like fragmentation of hard partons~\cite{K0sandLambda_PbPb, RAA2,jet_fragmentation_CMS}, which either are produced close to the surface of the medium~\cite{CMS:2021nhn} or escape from the QGP after in-medium interactions. Model calculations indicate that high-\pt hadrons are more likely to originate from narrow, hard fragmenting jets, which on average undergo fewer interactions with the QGP and thus lose less energy~\cite{Milhano:2015mng,Chesler:2015nqz,Rajagopal:2016uip}.

Within uncertainties, the measured \iaa shows no dependence on the trigger particle species for the considered $p_\mathrm{T,trig}$ range, where measurements of the nuclear modification factor \raa for different hadrons do not show any difference~\cite{STAR:2006uve,STAR:2011iap,RAA0,RAA1,RAA2}. Such an observation provides additional insight into parton energy loss mechanisms in strongly interacting  matter~\cite{Liu:2006sf}. Thus, the \iaa results together with \raa for identified hadrons can pose further constraints to theoretical models. The comparison of the \iaa measured from h--h correlations at \fivenn with the results obtained via $\pi^{0}$--h and h--h correlations in \PbPb collisions at \twosevensixnn, published in ~\cite{modification:SM,Jetlike_pion:SM}, is shown in the lower panels of Fig.~\ref{fig:all_IAA}.
It illustrates a good agreement between the results at the two different collision energies.

\begin{figure}[h!]
\begin{centering}  
\includegraphics[width=16cm]{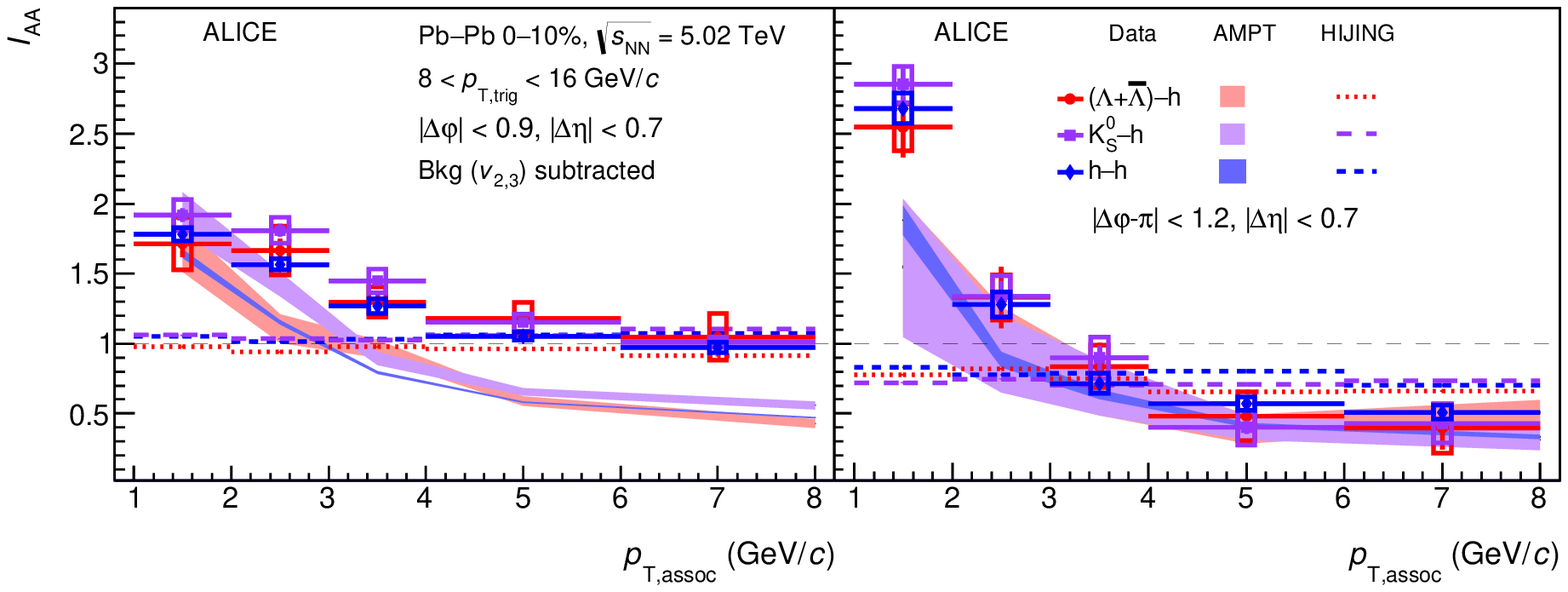}
\end{centering}
\caption{Per-trigger yield modification, \iaa, of \kzero--h (square points), ($\Lambda$+\almb)--h (circle points) and h--h (diamond points), on the near side (left) and away side (right), compared to model calculations from AMPT~\cite{AMPT_model,AMPT_model2} and HIJING~\cite{HIJING_model} event generators.}
\label{fig:all_IAA_model}
\end{figure}

The data are compared with theoretical models in Fig.~\ref{fig:all_IAA_model}, using the AMPT~\cite{AMPT_model,AMPT_model2} and HIJING~\cite{HIJING_model} event generators.
AMPT combines the initial particle distribution from HIJING with parton and hadron cascades with elastic scatterings for final-state interactions. String melting with a parton interaction cross section of 1.5 mb and parton recombination for hadronisation are used with parameters from~\cite{AMPT_model}.

On the away side (right-hand panel), the AMPT calculation reproduces the data within uncertainties except for the enhancement at the lowest $p_\mathrm{T,assoc}$, which is underestimated by the model.
On the near side, it qualitatively describes the enhancement for the two lowest $p_\mathrm{T,assoc}$ intervals but underestimates the measurements at high $p_\mathrm{T,assoc}$, where the predicted suppression is clearly not observed in the data. The disagreement in this region may be related to the fact that AMPT was found to overpredict the single-particle \RAA at the LHC energies compared to RHIC energies~\cite{AMPT_model3}.
The HIJING calculation, which includes the nuclear effects
of jet quenching and parton shadowing, shows no modification on the near side but a suppression on the away side almost independent of $p_\mathrm{T,assoc}$, which is different from the measured trend.

\newpage
\subsection{ \texorpdfstring{\Vdecay}{Lg}- to hadron-triggered yield ratios }

\begin{figure}[ht]
\includegraphics[width=16cm]{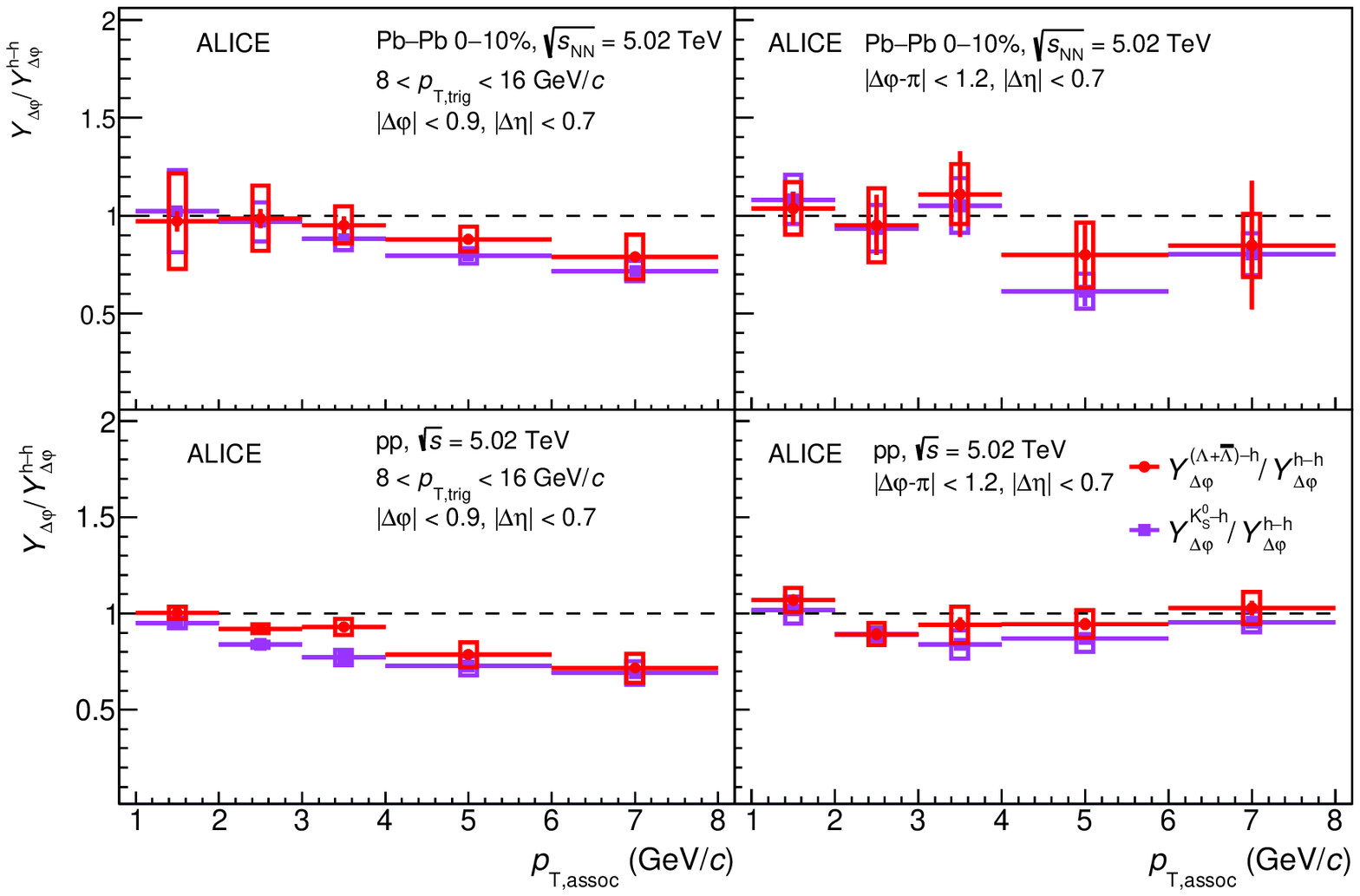}

\caption{Ratios of \Vdecay-triggered yields to charged-hadron triggered (h--h) ones, ${{Y}^{\kzero-\mathrm{h}}_{\Delta\varphi}}/{{Y}^{\mathrm{h-h}}_{\Delta\varphi}}$ (square points) and ${{Y}^{(\lmb+\almb)-\mathrm{h}}_{\Delta\varphi}}/{{Y}^{\mathrm{h-h}}_{\Delta\varphi}}$ (circle points), for \PbPb (top panels) and \pp (bottom panels) collisions in near side (left) and away side (right).}
\label{fig:yield_ratios}
\end{figure}

To investigate the differences between the \Vdecay-triggered and the unidentified h--h (a sample dominated by pions) yields, their ratios were calculated. In particular, the contribution of gluon jets may be different for \kzero and \lmb at high \pt, with baryons dominantly produced from gluon jets~\cite{Liu:2006sf}. These ratios as a function of $p_\mathrm{T,assoc}$ for both \pp and central \PbPb collisions are shown in Fig.~\ref{fig:yield_ratios}.
The results indicate that at low $p_\mathrm{T,assoc}$ (< 4 \GeVc ) on both the near and away sides, the differences between charged-hadron triggered yields and either \kzero or ($\Lambda$+\almb) triggered yields are small, and their ratios are compatible with unity at 1 < $p_\mathrm{T,assoc}$ < 2 \GeVc.
On the near side (left panels), their ratio goes below unity for $p_\mathrm{T,assoc}\gtrsim$ 3 \GeVc and it decreases with increasing $p_\mathrm{T,assoc}$, approaching 0.7 in the highest $p_\mathrm{T,assoc}$ interval of the measurement in both \pp and \PbPb collisions. This indicates a smaller amount of higher \pt particles in the jets triggered with a \Vdecay particle than in the jets triggered with an unidentified charged hadron, something previously observed by ALICE at $\sqrt{s}=$ 13 TeV in \pp collisions~\cite{hadron_correlations_pp}, and which was predicted by several theoretical models.

On the away side (right panels), the differences between \Vdecay and charged hadrons stay small and their ratio is consistent with unity within uncertainties in the whole $p_\mathrm{T,assoc}$ region analysed. This indicates that the back-to-back jets fragment independently and thus the away-side jet fragmentation is not affected by the trigger particle species. Furthermore, around 10\% difference of the near-side yields at intermediate \pt is observed between \kzero and ($\Lambda$+\almb) trigger particles in \pp collisions. However, this difference is not visible in \PbPb collisions within uncertainties. No difference between trigger \kzero and ($\Lambda$+\almb) is observed for the away-side yields for both collision systems.

\newpage

\newpage
\section{Summary}%
\label{section5}
Two-particle correlations have been measured as a function of the  azimuthal angle difference $\Delta\varphi$ at midrapidity in \pp and central \PbPb collisions at \fivenn, with \kzero, \lmb, and charged hadrons of transverse momenta $8 < p_\mathrm{T,trig} < 16$ $\mathrm{GeV}/c$ as trigger particles, and charged hadrons of transverse momenta $1 < p_\mathrm{T,assoc} < 8$ \GeVc as associated particles.

The per-trigger yields for $|\Delta\varphi|<0.9$ on the near side and $|\Delta\varphi-\pi|<1.2$ on the away side have been extracted. The ratios of the per-trigger yields of associated charged hadrons in \PbPb collisions with respect to \pp collisions, \iaa, have been calculated on the near and away sides. On the away side, the associated charged hadron yield in \PbPb collisions is suppressed relative to \pp collisions, resulting in \iaa $\approx$ 0.6 for $p_\mathrm{T,assoc}> 3$ \GeVc, while an enhancement increasing with decreasing momentum is observed, resulting in \iaa $\approx$ 2.7 at the lowest $p_\mathrm{T,assoc}$. On the near side, an enhancement up to \iaa $\approx$ 1.8 is observed at the lowest $p_\mathrm{T,assoc}$. The coincidence of the enhancement at low \pt on the near and away side as well as the away-side suppression at high \pt suggests a common underlying mechanism, perhaps related to the energy lost by high-momentum partons, which is recovered in the surrounding medium as required by local energy
conservation.

The \iaa of the yield associated with \kzero and ($\Lambda$+\almb) triggers has been compared with the one measured with inclusive charged hadrons as trigger particles, as well as with results obtained from $\pi^{0}-$h and h--h correlations at \twosevensixnn. They are in good agreement within uncertainties on both the near and away sides, where the measurements of \iaa show no dependence on the trigger particle species. This measured trigger-hadron independence of \iaa between strange and non-strange particles and between mesons and baryons provides new and important constraints for calculations of in-medium energy loss of high-\pt partons and their fragmentation properties, as well as the medium response to jets.

The measured \iaa is also compared to predictions from the AMPT and HIJING event generators: 
the AMPT calculation qualitatively describes the enhancement and suppression on the near and away sides, respectively. HIJING shows no modification on the near side and a suppression on the away side likely due to jet quenching but does not describe the~\pt dependence observed in the measurements.

Moreover, the ratios of the per-trigger yields, ${{Y}^{(\lmb+\almb)-\mathrm{h}}_{\Delta\varphi}}/{{Y}^{\mathrm{h-h}}_{\Delta\varphi}}$ and  ${{Y}^{\kzero-\mathrm{h}}_{\Delta\varphi}}/{{Y}^{\mathrm{h-h}}_{\Delta\varphi}}$, have been studied on the near and away sides in \pp and central \PbPb collisions. A decreasing trend of the ratio with increasing $p_\mathrm{T,assoc}$ is observed on the near side for both \PbPb and pp collisions, reaching values around 0.7 at the highest $p_\mathrm{T,assoc}$. 
This indicates a smaller amount of higher \pt particles in jets triggered with a \Vdecay particle than in jets triggered with an unidentified charged hadron.
In addition, a small difference between \kzero and \lmb triggered correlations observed on the near side in \pp collisions is not visible in \PbPb collisions within uncertainties. On the away side, the per-trigger yield ratio between \Vdecay and charged hadrons is consistent with unity within uncertainties in the whole $p_\mathrm{T,assoc}$ region analysed, proving that the recoiling jet is independent from the trigger particle species.

More precise and more differential studies are expected to be performed during Runs 3 and 4 of the LHC thanks to the upgraded ALICE detector~\cite{ALICE:upgrade}. By extending the measurements to various $p_\mathrm{T,trig}$-intervals of (multi)strange particles in different centrality classes, hot-medium effects can be further investigated on the modification of the associated particle yields and correlation widths. This will provide additional important information on understanding of the jet-medium interaction in ultra-relativistic heavy-ion collisions.

%%%%%%%%%%%%%%%%%%%%%%%%%%%%%%%%
% end main text 
%%%%%%%%%%%%%%%%%%%%%%%%%%%%%%%%

%%%%% acknowledgements - handled by EB chairs 
\newenvironment{acknowledgement}{\relax}{\relax}
\begin{acknowledgement}
\section*{Acknowledgements}
% add specific acknowledgements here 
% ...but please don't remove the line below: funding agencies
% will be acknowledged with a custom tex file handled by EB chairs after Collab Round 2
% Version: 2022-10-10

The ALICE Collaboration would like to thank all its engineers and technicians for their invaluable contributions to the construction of the experiment and the CERN accelerator teams for the outstanding performance of the LHC complex.
The ALICE Collaboration gratefully acknowledges the resources and support provided by all Grid centres and the Worldwide LHC Computing Grid (WLCG) collaboration.
The ALICE Collaboration acknowledges the following funding agencies for their support in building and running the ALICE detector:
A. I. Alikhanyan National Science Laboratory (Yerevan Physics Institute) Foundation (ANSL), State Committee of Science and World Federation of Scientists (WFS), Armenia;
Austrian Academy of Sciences, Austrian Science Fund (FWF): [M 2467-N36] and Nationalstiftung f\"{u}r Forschung, Technologie und Entwicklung, Austria;
Ministry of Communications and High Technologies, National Nuclear Research Center, Azerbaijan;
Conselho Nacional de Desenvolvimento Cient\'{\i}fico e Tecnol\'{o}gico (CNPq), Financiadora de Estudos e Projetos (Finep), Funda\c{c}\~{a}o de Amparo \`{a} Pesquisa do Estado de S\~{a}o Paulo (FAPESP) and Universidade Federal do Rio Grande do Sul (UFRGS), Brazil;
Bulgarian Ministry of Education and Science, within the National Roadmap for Research Infrastructures 2020¿2027 (object CERN), Bulgaria;
Ministry of Education of China (MOEC) , Ministry of Science \& Technology of China (MSTC) and National Natural Science Foundation of China (NSFC), China;
Ministry of Science and Education and Croatian Science Foundation, Croatia;
Centro de Aplicaciones Tecnol\'{o}gicas y Desarrollo Nuclear (CEADEN), Cubaenerg\'{\i}a, Cuba;
Ministry of Education, Youth and Sports of the Czech Republic, Czech Republic;
The Danish Council for Independent Research | Natural Sciences, the VILLUM FONDEN and Danish National Research Foundation (DNRF), Denmark;
Helsinki Institute of Physics (HIP), Finland;
Commissariat \`{a} l'Energie Atomique (CEA) and Institut National de Physique Nucl\'{e}aire et de Physique des Particules (IN2P3) and Centre National de la Recherche Scientifique (CNRS), France;
Bundesministerium f\"{u}r Bildung und Forschung (BMBF) and GSI Helmholtzzentrum f\"{u}r Schwerionenforschung GmbH, Germany;
General Secretariat for Research and Technology, Ministry of Education, Research and Religions, Greece;
National Research, Development and Innovation Office, Hungary;
Department of Atomic Energy Government of India (DAE), Department of Science and Technology, Government of India (DST), University Grants Commission, Government of India (UGC) and Council of Scientific and Industrial Research (CSIR), India;
National Research and Innovation Agency - BRIN, Indonesia;
Istituto Nazionale di Fisica Nucleare (INFN), Italy;
Japanese Ministry of Education, Culture, Sports, Science and Technology (MEXT) and Japan Society for the Promotion of Science (JSPS) KAKENHI, Japan;
Consejo Nacional de Ciencia (CONACYT) y Tecnolog\'{i}a, through Fondo de Cooperaci\'{o}n Internacional en Ciencia y Tecnolog\'{i}a (FONCICYT) and Direcci\'{o}n General de Asuntos del Personal Academico (DGAPA), Mexico;
Nederlandse Organisatie voor Wetenschappelijk Onderzoek (NWO), Netherlands;
The Research Council of Norway, Norway;
Commission on Science and Technology for Sustainable Development in the South (COMSATS), Pakistan;
Pontificia Universidad Cat\'{o}lica del Per\'{u}, Peru;
Ministry of Education and Science, National Science Centre and WUT ID-UB, Poland;
Korea Institute of Science and Technology Information and National Research Foundation of Korea (NRF), Republic of Korea;
Ministry of Education and Scientific Research, Institute of Atomic Physics, Ministry of Research and Innovation and Institute of Atomic Physics and University Politehnica of Bucharest, Romania;
Ministry of Education, Science, Research and Sport of the Slovak Republic, Slovakia;
National Research Foundation of South Africa, South Africa;
Swedish Research Council (VR) and Knut \& Alice Wallenberg Foundation (KAW), Sweden;
European Organization for Nuclear Research, Switzerland;
Suranaree University of Technology (SUT), National Science and Technology Development Agency (NSTDA), Thailand Science Research and Innovation (TSRI) and National Science, Research and Innovation Fund (NSRF), Thailand;
Turkish Energy, Nuclear and Mineral Research Agency (TENMAK), Turkey;
National Academy of  Sciences of Ukraine, Ukraine;
Science and Technology Facilities Council (STFC), United Kingdom;
National Science Foundation of the United States of America (NSF) and United States Department of Energy, Office of Nuclear Physics (DOE NP), United States of America.
In addition, individual groups or members have received support from:
Marie Sk\l{}odowska Curie, European Research Council, Strong 2020 - Horizon 2020 (grant nos. 950692, 824093, 896850), European Union;
Academy of Finland (Center of Excellence in Quark Matter) (grant nos. 346327, 346328), Finland;
Programa de Apoyos para la Superaci\'{o}n del Personal Acad\'{e}mico, UNAM, Mexico.

\end{acknowledgement}

%%%%%%%% Bibliography 
\bibliographystyle{utphys}
\bibliography{bibliography}

%%%%%%%%%%%%%%%%%%%%%%%%%%%%%%%%
% Appendices: yours (if any) + authorlist
%%%%%%%%%%%%%%%%%%%%%%%%%%%%%%%%
\newpage
\appendix
%%%----------------------------------------------------------
\section{Appendix}
\label{app:appendix}
%\input{appendix.tex}

%\appendix
%\input{} % put your appendices here (if any)

%\textbf{A   Appendix A}\\

\vspace{0.2cm}

\textbf{A.1 \hspace{0.1cm}   \Vdecay to hadron-triggered yield ratios compared to results at $\sqrt{s}=$ 13 TeV}

\begin{figure}[ht]
%\centering
%\begin{array}{ccc}
\includegraphics[width=16cm]{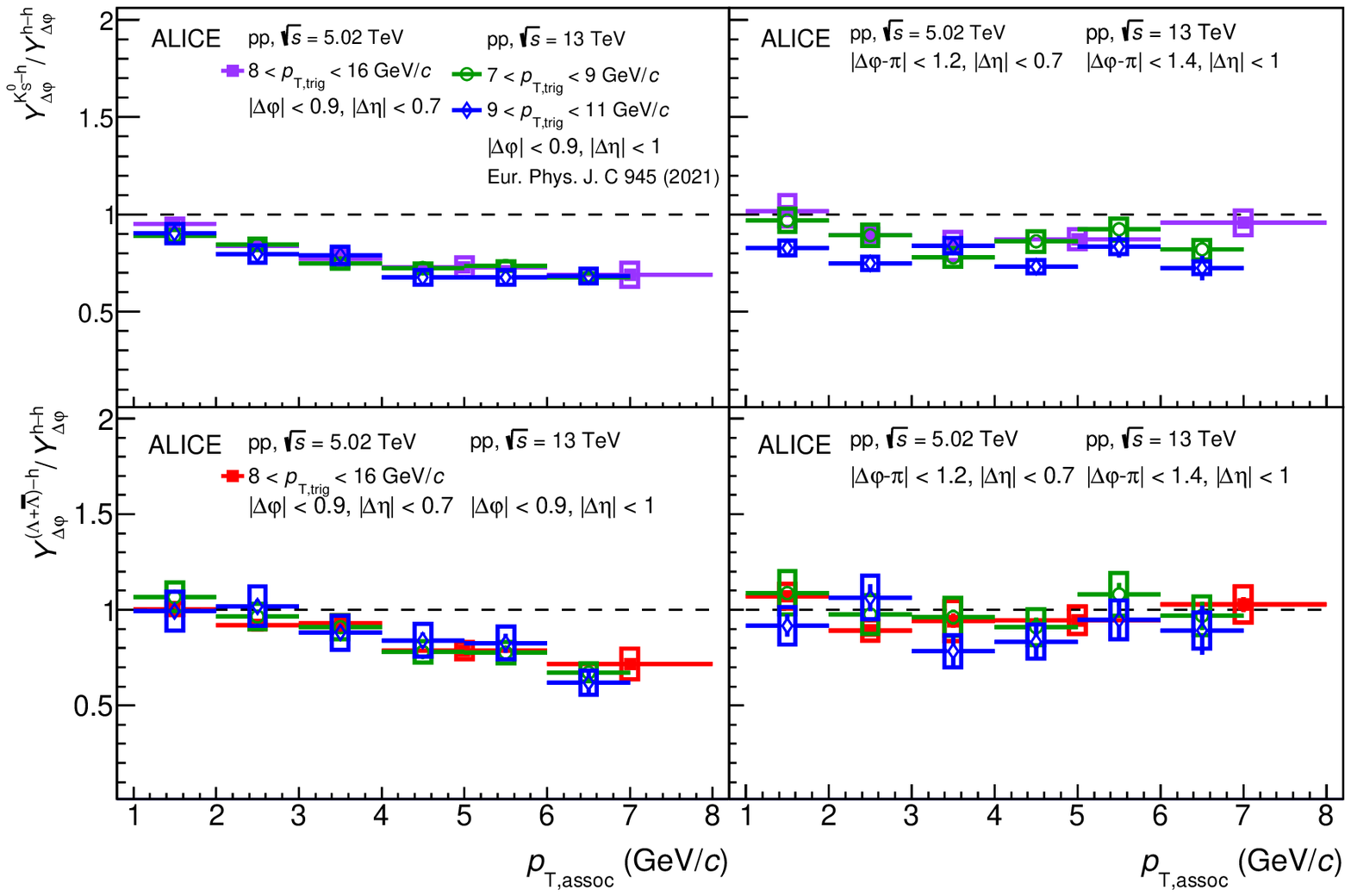}

%\end{array}$
\caption{Ratios of \Vdecay-triggered yields to charged-hadron (h--h) yields, ${{Y}^{\kzero-\mathrm{h}}_{\Delta\varphi}}/{{Y}^{\mathrm{h-h}}_{\Delta\varphi}}$ (top panels) and ${{Y}^{(\lmb+\almb)-\mathrm{h}}_{\Delta\varphi}}/{{Y}^{\mathrm{h-h}}_{\Delta\varphi}}$ (bottom panels), for \pp collisions  on the near side (left) and away side (right) from this measurement compared with published results at higher collision energy~\cite{hadron_correlations_pp}.}
\label{fig:yield_ratios_comparision}

\end{figure}

The yield ratios on the near and away sides, ${{Y}^{\kzero-\mathrm{h}}_{\Delta\varphi}}/{{Y}^{\mathrm{h-h}}_{\Delta\varphi}}$ and  ${{Y}^{(\lmb+\almb)-\mathrm{h}}_{\Delta\varphi}}/{{Y}^{\mathrm{h-h}}_{\Delta\varphi}}$, in \pp collisions at $\sqrt{s}= 5.02$ TeV are compared in Fig.~\ref{fig:yield_ratios_comparision} to the corresponding results from \pp collisions at $\sqrts$ =13 TeV~\cite{hadron_correlations_pp} with slight differences in $p_\mathrm{T,trig}$ intervals, $\deta$ and $\dphi$ ranges. The results show that the ratio do not depend on $\sqrt{s}$ and $p_\mathrm{T,trig}$ intervals within uncertainties. This is consistent with the fragmentation function being weakly dependent on collision energy in the LHC domain.

\clearpage

%%%%% Authorlist - please do not touch: handled by EB chairs 
\section{The ALICE Collaboration}
\label{app:collab}
% ALICE Collaboration author list for 2022-10-10
\begin{flushleft} 
\small

S.~Acharya\,\orcidlink{0000-0002-9213-5329}\,$^{\rm 125}$, 
D.~Adamov\'{a}\,\orcidlink{0000-0002-0504-7428}\,$^{\rm 86}$, 
A.~Adler$^{\rm 69}$, 
G.~Aglieri Rinella\,\orcidlink{0000-0002-9611-3696}\,$^{\rm 32}$, 
M.~Agnello\,\orcidlink{0000-0002-0760-5075}\,$^{\rm 29}$, 
N.~Agrawal\,\orcidlink{0000-0003-0348-9836}\,$^{\rm 50}$, 
Z.~Ahammed\,\orcidlink{0000-0001-5241-7412}\,$^{\rm 132}$, 
S.~Ahmad\,\orcidlink{0000-0003-0497-5705}\,$^{\rm 15}$, 
S.U.~Ahn\,\orcidlink{0000-0001-8847-489X}\,$^{\rm 70}$, 
I.~Ahuja\,\orcidlink{0000-0002-4417-1392}\,$^{\rm 37}$
% ...

\section*{Affiliation Notes}

$^{\rm I}$ Deceased\\
$^{\rm II}$ Also at: Max-Planck-Institut f\"{u}r Physik, Munich, Germany\\
$^{\rm III}$ Also at: Italian National Agency for New Technologies, Energy and Sustainable Economic Development (ENEA), Bologna, Italy\\
$^{\rm IV}$ Also at: Dipartimento DET del Politecnico di Torino, Turin, Italy\\
$^{\rm V}$ Also at: Department of Applied Physics, Aligarh Muslim University, Aligarh, India\\
$^{\rm VI}$ Also at: Institute of Theoretical Physics, University of Wroclaw, Poland\\
$^{\rm VII}$ Also at: An institution covered by a cooperation agreement with CERN\\

\section*{Collaboration Institutes}

$^{1}$ A.I. Alikhanyan National Science Laboratory (Yerevan Physics Institute) Foundation, Yerevan, Armenia\\
$^{2}$ AGH University of Science and Technology, Cracow, Poland\\
$^{3}$ Bogolyubov Institute for Theoretical Physics, National Academy of Sciences of Ukraine, Kiev, Ukraine\\
$^{4}$ Bose Institute, Department of Physics  and Centre for Astroparticle Physics and Space Science (CAPSS), Kolkata, India\\
$^{5}$ California Polytechnic State University, San Luis Obispo, California, United States\\
$^{6}$ Central China Normal University, Wuhan, China\\
$^{7}$ Centro de Aplicaciones Tecnol\'{o}gicas y Desarrollo Nuclear (CEADEN), Havana, Cuba\\
$^{8}$ Centro de Investigaci\'{o}n y de Estudios Avanzados (CINVESTAV), Mexico City and M\'{e}rida, Mexico\\
$^{9}$ Chicago State University, Chicago, Illinois, United States\\
$^{10}$ China Institute of Atomic Energy, Beijing, China\\
$^{11}$ Chungbuk National University, Cheongju, Republic of Korea\\
$^{12}$ Comenius University Bratislava, Faculty of Mathematics, Physics and Informatics, Bratislava, Slovak Republic\\
$^{13}$ COMSATS University Islamabad, Islamabad, Pakistan\\
$^{14}$ Creighton University, Omaha, Nebraska, United States\\
$^{15}$ Department of Physics, Aligarh Muslim University, Aligarh, India\\
$^{16}$ Department of Physics, Pusan National University, Pusan, Republic of Korea\\
$^{17}$ Department of Physics, Sejong University, Seoul, Republic of Korea\\
$^{18}$ Department of Physics, University of California, Berkeley, California, United States\\
$^{19}$ Department of Physics, University of Oslo, Oslo, Norway\\
$^{20}$ Department of Physics and Technology, University of Bergen, Bergen, Norway\\
$^{21}$ Dipartimento di Fisica, Universit\`{a} di Pavia, Pavia, Italy\\
$^{22}$ Dipartimento di Fisica dell'Universit\`{a} and Sezione INFN, Cagliari, Italy\\
$^{23}$ Dipartimento di Fisica dell'Universit\`{a} and Sezione INFN, Trieste, Italy\\
$^{24}$ Dipartimento di Fisica dell'Universit\`{a} and Sezione INFN, Turin, Italy\\
$^{25}$ Dipartimento di Fisica e Astronomia dell'Universit\`{a} and Sezione INFN, Bologna, Italy\\
$^{26}$ Dipartimento di Fisica e Astronomia dell'Universit\`{a} and Sezione INFN, Catania, Italy\\
$^{27}$ Dipartimento di Fisica e Astronomia dell'Universit\`{a} and Sezione INFN, Padova, Italy\\
$^{28}$ Dipartimento di Fisica `E.R.~Caianiello' dell'Universit\`{a} and Gruppo Collegato INFN, Salerno, Italy\\
$^{29}$ Dipartimento DISAT del Politecnico and Sezione INFN, Turin, Italy\\
$^{30}$ Dipartimento di Scienze MIFT, Universit\`{a} di Messina, Messina, Italy\\
$^{31}$ Dipartimento Interateneo di Fisica `M.~Merlin' and Sezione INFN, Bari, Italy\\
$^{32}$ European Organization for Nuclear Research (CERN), Geneva, Switzerland\\
$^{33}$ Faculty of Electrical Engineering, Mechanical Engineering and Naval Architecture, University of Split, Split, Croatia\\
$^{34}$ Faculty of Engineering and Science, Western Norway University of Applied Sciences, Bergen, Norway\\
$^{35}$ Faculty of Nuclear Sciences and Physical Engineering, Czech Technical University in Prague, Prague, Czech Republic\\
$^{36}$ Faculty of Physics, Sofia University, Sofia, Bulgaria\\
$^{37}$ Faculty of Science, P.J.~\v{S}af\'{a}rik University, Ko\v{s}ice, Slovak Republic\\
$^{38}$ Frankfurt Institute for Advanced Studies, Johann Wolfgang Goethe-Universit\"{a}t Frankfurt, Frankfurt, Germany\\
$^{39}$ Fudan University, Shanghai, China\\
$^{40}$ Gangneung-Wonju National University, Gangneung, Republic of Korea\\
$^{41}$ Gauhati University, Department of Physics, Guwahati, India\\
$^{42}$ Helmholtz-Institut f\"{u}r Strahlen- und Kernphysik, Rheinische Friedrich-Wilhelms-Universit\"{a}t Bonn, Bonn, Germany\\
$^{43}$ Helsinki Institute of Physics (HIP), Helsinki, Finland\\
$^{44}$ High Energy Physics Group,  Universidad Aut\'{o}noma de Puebla, Puebla, Mexico\\
$^{45}$ Horia Hulubei National Institute of Physics and Nuclear Engineering, Bucharest, Romania\\
$^{46}$ Indian Institute of Technology Bombay (IIT), Mumbai, India\\
$^{47}$ Indian Institute of Technology Indore, Indore, India\\
$^{48}$ INFN, Laboratori Nazionali di Frascati, Frascati, Italy\\
$^{49}$ INFN, Sezione di Bari, Bari, Italy\\
$^{50}$ INFN, Sezione di Bologna, Bologna, Italy\\
$^{51}$ INFN, Sezione di Cagliari, Cagliari, Italy\\
$^{52}$ INFN, Sezione di Catania, Catania, Italy\\
$^{53}$ INFN, Sezione di Padova, Padova, Italy\\
$^{54}$ INFN, Sezione di Pavia, Pavia, Italy\\
$^{55}$ INFN, Sezione di Torino, Turin, Italy\\
$^{56}$ INFN, Sezione di Trieste, Trieste, Italy\\
$^{57}$ Inha University, Incheon, Republic of Korea\\
$^{58}$ Institute for Gravitational and Subatomic Physics (GRASP), Utrecht University/Nikhef, Utrecht, Netherlands\\
$^{59}$ Institute of Experimental Physics, Slovak Academy of Sciences, Ko\v{s}ice, Slovak Republic\\
$^{60}$ Institute of Physics, Homi Bhabha National Institute, Bhubaneswar, India\\
$^{61}$ Institute of Physics of the Czech Academy of Sciences, Prague, Czech Republic\\
$^{62}$ Institute of Space Science (ISS), Bucharest, Romania\\
$^{63}$ Institut f\"{u}r Kernphysik, Johann Wolfgang Goethe-Universit\"{a}t Frankfurt, Frankfurt, Germany\\
$^{64}$ Instituto de Ciencias Nucleares, Universidad Nacional Aut\'{o}noma de M\'{e}xico, Mexico City, Mexico\\
$^{65}$ Instituto de F\'{i}sica, Universidade Federal do Rio Grande do Sul (UFRGS), Porto Alegre, Brazil\\
$^{66}$ Instituto de F\'{\i}sica, Universidad Nacional Aut\'{o}noma de M\'{e}xico, Mexico City, Mexico\\
$^{67}$ iThemba LABS, National Research Foundation, Somerset West, South Africa\\
$^{68}$ Jeonbuk National University, Jeonju, Republic of Korea\\
$^{69}$ Johann-Wolfgang-Goethe Universit\"{a}t Frankfurt Institut f\"{u}r Informatik, Fachbereich Informatik und Mathematik, Frankfurt, Germany\\
$^{70}$ Korea Institute of Science and Technology Information, Daejeon, Republic of Korea\\
$^{71}$ KTO Karatay University, Konya, Turkey\\
$^{72}$ Laboratoire de Physique des 2 Infinis, Ir\`{e}ne Joliot-Curie, Orsay, France\\
$^{73}$ Laboratoire de Physique Subatomique et de Cosmologie, Universit\'{e} Grenoble-Alpes, CNRS-IN2P3, Grenoble, France\\
$^{74}$ Lawrence Berkeley National Laboratory, Berkeley, California, United States\\
$^{75}$ Lund University Department of Physics, Division of Particle Physics, Lund, Sweden\\
$^{76}$ Nagasaki Institute of Applied Science, Nagasaki, Japan\\
$^{77}$ Nara Women{'}s University (NWU), Nara, Japan\\
$^{78}$ National and Kapodistrian University of Athens, School of Science, Department of Physics , Athens, Greece\\
$^{79}$ National Centre for Nuclear Research, Warsaw, Poland\\
$^{80}$ National Institute of Science Education and Research, Homi Bhabha National Institute, Jatni, India\\
$^{81}$ National Nuclear Research Center, Baku, Azerbaijan\\
$^{82}$ National Research and Innovation Agency - BRIN, Jakarta, Indonesia\\
$^{83}$ Niels Bohr Institute, University of Copenhagen, Copenhagen, Denmark\\
$^{84}$ Nikhef, National institute for subatomic physics, Amsterdam, Netherlands\\
$^{85}$ Nuclear Physics Group, STFC Daresbury Laboratory, Daresbury, United Kingdom\\
$^{86}$ Nuclear Physics Institute of the Czech Academy of Sciences, Husinec-\v{R}e\v{z}, Czech Republic\\
$^{87}$ Oak Ridge National Laboratory, Oak Ridge, Tennessee, United States\\
$^{88}$ Ohio State University, Columbus, Ohio, United States\\
$^{89}$ Physics department, Faculty of science, University of Zagreb, Zagreb, Croatia\\
$^{90}$ Physics Department, Panjab University, Chandigarh, India\\
$^{91}$ Physics Department, University of Jammu, Jammu, India\\
$^{92}$ Physics Program and International Institute for Sustainability with Knotted Chiral Meta Matter (SKCM2), Hiroshima University, Hiroshima, Japan\\
$^{93}$ Physikalisches Institut, Eberhard-Karls-Universit\"{a}t T\"{u}bingen, T\"{u}bingen, Germany\\
$^{94}$ Physikalisches Institut, Ruprecht-Karls-Universit\"{a}t Heidelberg, Heidelberg, Germany\\
$^{95}$ Physik Department, Technische Universit\"{a}t M\"{u}nchen, Munich, Germany\\
$^{96}$ Politecnico di Bari and Sezione INFN, Bari, Italy\\
$^{97}$ Research Division and ExtreMe Matter Institute EMMI, GSI Helmholtzzentrum f\"ur Schwerionenforschung GmbH, Darmstadt, Germany\\
$^{98}$ Saga University, Saga, Japan\\
$^{99}$ Saha Institute of Nuclear Physics, Homi Bhabha National Institute, Kolkata, India\\
$^{100}$ School of Physics and Astronomy, University of Birmingham, Birmingham, United Kingdom\\
$^{101}$ Secci\'{o}n F\'{\i}sica, Departamento de Ciencias, Pontificia Universidad Cat\'{o}lica del Per\'{u}, Lima, Peru\\
$^{102}$ Stefan Meyer Institut f\"{u}r Subatomare Physik (SMI), Vienna, Austria\\
$^{103}$ SUBATECH, IMT Atlantique, Nantes Universit\'{e}, CNRS-IN2P3, Nantes, France\\
$^{104}$ Sungkyunkwan University, Suwon City, Republic of Korea\\
$^{105}$ Suranaree University of Technology, Nakhon Ratchasima, Thailand\\
$^{106}$ Technical University of Ko\v{s}ice, Ko\v{s}ice, Slovak Republic\\
$^{107}$ The Henryk Niewodniczanski Institute of Nuclear Physics, Polish Academy of Sciences, Cracow, Poland\\
$^{108}$ The University of Texas at Austin, Austin, Texas, United States\\
$^{109}$ Universidad Aut\'{o}noma de Sinaloa, Culiac\'{a}n, Mexico\\
$^{110}$ Universidade de S\~{a}o Paulo (USP), S\~{a}o Paulo, Brazil\\
$^{111}$ Universidade Estadual de Campinas (UNICAMP), Campinas, Brazil\\
$^{112}$ Universidade Federal do ABC, Santo Andre, Brazil\\
$^{113}$ University of Cape Town, Cape Town, South Africa\\
$^{114}$ University of Houston, Houston, Texas, United States\\
$^{115}$ University of Jyv\"{a}skyl\"{a}, Jyv\"{a}skyl\"{a}, Finland\\
$^{116}$ University of Kansas, Lawrence, Kansas, United States\\
$^{117}$ University of Liverpool, Liverpool, United Kingdom\\
$^{118}$ University of Science and Technology of China, Hefei, China\\
$^{119}$ University of South-Eastern Norway, Kongsberg, Norway\\
$^{120}$ University of Tennessee, Knoxville, Tennessee, United States\\
$^{121}$ University of the Witwatersrand, Johannesburg, South Africa\\
$^{122}$ University of Tokyo, Tokyo, Japan\\
$^{123}$ University of Tsukuba, Tsukuba, Japan\\
$^{124}$ University Politehnica of Bucharest, Bucharest, Romania\\
$^{125}$ Universit\'{e} Clermont Auvergne, CNRS/IN2P3, LPC, Clermont-Ferrand, France\\
$^{126}$ Universit\'{e} de Lyon, CNRS/IN2P3, Institut de Physique des 2 Infinis de Lyon, Lyon, France\\
$^{127}$ Universit\'{e} de Strasbourg, CNRS, IPHC UMR 7178, F-67000 Strasbourg, France, Strasbourg, France\\
$^{128}$ Universit\'{e} Paris-Saclay Centre d'Etudes de Saclay (CEA), IRFU, D\'{e}partment de Physique Nucl\'{e}aire (DPhN), Saclay, France\\
$^{129}$ Universit\`{a} degli Studi di Foggia, Foggia, Italy\\
$^{130}$ Universit\`{a} del Piemonte Orientale, Vercelli, Italy\\
$^{131}$ Universit\`{a} di Brescia, Brescia, Italy\\
$^{132}$ Variable Energy Cyclotron Centre, Homi Bhabha National Institute, Kolkata, India\\
$^{133}$ Warsaw University of Technology, Warsaw, Poland\\
$^{134}$ Wayne State University, Detroit, Michigan, United States\\
$^{135}$ Westf\"{a}lische Wilhelms-Universit\"{a}t M\"{u}nster, Institut f\"{u}r Kernphysik, M\"{u}nster, Germany\\
$^{136}$ Wigner Research Centre for Physics, Budapest, Hungary\\
$^{137}$ Yale University, New Haven, Connecticut, United States\\
$^{138}$ Yonsei University, Seoul, Republic of Korea\\
$^{139}$  Zentrum  f\"{u}r Technologie und Transfer (ZTT), Worms, Germany\\
$^{140}$ Affiliated with an institute covered by a cooperation agreement with CERN\\
$^{141}$ Affiliated with an international laboratory covered by a cooperation agreement with CERN.\\

\end{flushleft} 

\end{document}